\documentclass[preprint]{elsarticle}
\usepackage{hyperref}
\usepackage[dvipsnames]{xcolor}
\usepackage{longtable}
\usepackage{pstool}
\usepackage{geometry}
\biboptions{numbers,sort&compress}
\usepackage{tikz}

\graphicspath{ {./figures/} {./fig/} }
\usepackage{booktabs, siunitx, array, threeparttable}
\usepackage{makecell}
\usepackage{nomencl}
\makenomenclature
\usepackage{etoolbox}
\renewcommand\nomgroup[1]{%
  \ifstrequal{#1}{S}{\item[\itshape Subscripts]}{%
  \ifstrequal{#1}{G}{\item[\itshape Greek symbols]}{}}}
\usepackage{caption}
\usepackage{subcaption}
\usepackage{amsmath}
\usepackage{amssymb}
\usepackage{xcolor}
\usepackage{lscape}
\usepackage{booktabs}% http://ctan.org/pkg/booktabs

\usepackage{multirow}
\usepackage{amsmath}
\usepackage{xcolor}
\usepackage{textcomp}
\usepackage{color,soul}

\usepackage{ulem}
\usepackage{url}
\usepackage{graphicx, tabularx}
\usepackage{float}
\definecolor{darkgreen}{rgb}{0.0, 0.5, 0.0}
\definecolor{bblue}{rgb}{0.0, 0.439, 0.753}
\definecolor{viridian}{rgb}{0.25, 0.51, 0.43}
\definecolor{Al000}{HTML}{95CEFC}
\definecolor{Al001}{HTML}{7EAFD6}
\definecolor{Al010}{HTML}{688FB0}
\definecolor{Al100}{HTML}{425B70}
\def\Nu{\textup{Nu}}
\def\Re{\textup{Re}}
\def\NTU{\textup{NTU}}

\def\M{\textup{M}}
\begin{document}
\makeatletter
\def\ps@pprintTitle{%
  \let\@oddhead\@empty
  \let\@evenhead\@empty
  \let\@oddfoot\@empty
  \let\@evenfoot\@empty
}
\makeatother
\begin{frontmatter}
\title{Optimal Sizing and Material Choice for Additively Manufactured Compact Plate Heat Exchangers}
\author[1,3]{M.~Basaran}\corref{cor1}
\ead{mehmet.basaran@kuleuven.be}
\cortext[cor1]{Corresponding Author}
\author[2]{F.~Rogiers}
\ead{frederik.rogiers@kuleuven.be}
\author[2,3]{M.~Baelmans}
\ead{martine.baelmans@kuleuven.be}
\author[1,3]{M.~Blommaert}
\ead{maarten.blommaert@kuleuven.be}
\address[1]{KU Leuven, Department of Mechanical Engineering, Faculty of Engineering Technology, B-2440 Geel, Belgium}
\address[2]{KU Leuven, Department of Mechanical Engineering, Division of Applied Mechanics and Energy Conversion (TME), B-3001 Leuven, Belgium}
\address[3]{EnergyVille, Thor Park, B-3600 Genk, Belgium}
%%%%%%%%%%%%%%%%
\nomenclature[Aa]{$A$}{heat transfer surface area [m$^2$]}
\nomenclature[Aax]{$A_{ax}$}{heat transfer surface area in the axial direction [m$^2$]}
\nomenclature[Acp]{$c_p$}{fluid specific heat at constant pressure [J/kg$\cdot$K]}
\nomenclature[Ad]{$D$}{plate spacing [m]}
\nomenclature[Adstar]{$D^\ast$}{nondimensional plate spacing, $D/t_\text{ref}$ [-]}
\nomenclature[Adcal]{$\mathcal{D}$}{thermal power density, $\dot{q}/\left( H L W\right)$ [W/m$^3$]}
\nomenclature[Af]{$f$}{Fanning friction factor [-]}
\nomenclature[Ah]{$H$}{total heat exchanger height [m]}
\nomenclature[Ak]{$k$}{fluid thermal conductivity [W/m$\cdot$K]}
\nomenclature[Akw]{$k_w$}{plate thermal conductivity [W/m$\cdot$K]}
\nomenclature[Al]{$L$}{plate length, channel length [m]}
\nomenclature[Alstar]{$L^\ast$}{nondimensional plate length, $L/t_\text{ref}$ [-]}
\nomenclature[Am]{$\dot{m}$}{mass flow rate at each side [kg/s]}
\nomenclature[Amprime]{$\dot{m}^{\prime}_\text{single}$}{mass flow rate per unit width for a single channel, $\dot{m}/\left(nW\right)$ [kg/s$\cdot$m]}
\nomenclature[AM]{$\M$}{axial conduction parameter, $UA_{ax}/\left( \dot{m} c_p\right)$ [-]}
\nomenclature[An]{$n$}{number of hot side or cold side channels [-]}
\nomenclature[Antu]{$\NTU$}{number of transfer units, $UA/\left( \dot{m} c_p\right)$ [-]}
\nomenclature[Ap]{$\Delta P$}{pressure drop between inlet and outlet [Pa]}
\nomenclature[Aqdot]{$\dot{q}$}{heat transfer rate [W]}
\nomenclature[Aqstar]{$Q^{\diamond}$}{nondimensional power density, Eq~\ref{NPD} [-]}
\nomenclature[ARe]{$\Re$}{Reynolds number based on 2D, $\rho V2D/\mu$ [-]}
\nomenclature[At]{$t$}{plate thickness [m]}
\nomenclature[Atstar]{$t^\ast$}{nondimensional plate thickness, $t/t_\text{ref}$ [-]}
\nomenclature[Atz]{$\Delta T$}{hot to cold side inlet temperature difference [K]}
\nomenclature[AU]{$U$}{heat transfer coefficient [W/m$^2$$\cdot$K]} 
\nomenclature[AV]{$V$}{fluid bulk velocity in parallel-plate channel [m/s]}
\nomenclature[AW]{$W$}{plate width, channel width [m]}
\nomenclature[AIF]{IF}{Improvement Factor [-]}
\nomenclature[ANu]{$\Nu$}{mean Nusselt number, h2D/k [-]}
\nomenclature[Galpha]{$\alpha$}{fluid thermal diffusivity, $k/\left(\rho c_p\right)$ [m$^2$/s]}
\nomenclature[Ggamma]{$\gamma$}{plate thickness-to-spacing ratio constant, $t^\ast/D^\ast$ [-]}
\nomenclature[Gpi]{$\Pi$}{dimensionless scaling group $\Delta P_{\text{ref}} t_{\text{ref}}^2/\left(\alpha \mu\right)$ [-]}
\nomenclature[Gpsi]{$\psi$}{ratio of the actual pressure drop to the reference value, $\Delta P/\Delta P_{\text{ref}}$ [-]}
\nomenclature[Grho]{$\rho$}{fluid density [kg/m$^3$]}
\nomenclature[Geps]{$\varepsilon$}{heat exchanger effectiveness [-]}
\nomenclature[Gmu]{$\mu$}{fluid dynamic viscosity [Pa$\cdot$s]}
\nomenclature[Sd]{d}{design value}
\nomenclature[Smin]{min}{minimum allowable value}
\nomenclature[Sref]{ref}{reference value}
%%%%%%%%%%%%%%%%
\begin{abstract}
Advances in additive manufacturing (AM) enable new opportunities to design compact heat exchangers (cHEXs) by leveraging flexible geometries to improve energy and material efficiency. However, it is well-known that reducing size in counterflow cHEXs can degrade effectiveness due to axial heat conduction through the solid material, which depends strongly on material thermal conductivity and wall thickness. Understanding this interaction between fundamental heat transfer mechanisms and manufacturing constraints is essential for designing next-generation compact thermal systems with the aspiration of taking full advantage of AM's shaping flexibility. This study investigates how material selection and AM thin-wall limitations influence the maximum achievable power density in compact plate heat exchangers. An optimization framework evaluates six materials: plastic, austenitic steel, Al\textsubscript{2}O\textsubscript{3}, AlN, aluminum, and copper. The evaluations are performed under fixed pressure drop and effectiveness, while accounting for AM-specific thickness constraints and a lower limit on plate spacing to address fouling risks. Results show that copper consistently yields the lowest power density, despite having the highest thermal conductivity of the materials considered. In contrast, plastic achieves the highest power density across the large majority of optimization scenarios. Without manufacturing or fouling constraints, plastic outperforms the baseline steel design by a factor of 962. With a uniform plate thickness of 0.5\,mm, it remains the top performer with 21.6 times the power density of copper. Adding a fouling constraint narrows the performance gap, making austenitic steel and plastic comparable. When material-specific thicknesses are applied, plastic again leads in compactness with 7.5 times the power density of copper, due to its superior thin-wall manufacturability. These findings highlight that AM constraints substantially affect cHEX compactness, and that lower-conductivity materials like plastic and austenitic steel can outperform metals like copper in power-dense heat exchanger designs. The results demonstrate how manufacturing limits, material properties, and heat transfer fundamentals jointly determine achievable compactness in AM-based heat exchangers.
\end{abstract}
\end{frontmatter}
%
% \linenumbers
%%%
\begingroup
\small
\printnomenclature[1cm]
\endgroup
%%%
\section{Introduction}\label{sec:introduction}

%%%%%%%%%%%%%%%%% Growing demands on heat exchanger design %%%%%%%%%%%%%%%%%

Heat exchangers are crucial components in various engineering fields, including the automotive and aerospace industries, microelectronics cooling, and energy conversion systems. Increasing demands for compact and lightweight solutions, particularly in mobile and decentralized energy applications, have amplified the need for high-performance compact heat exchangers (cHEXs)~\cite{Wang2024,Broatch2024}. As heat transfer applications move toward higher power density and smaller scales, connecting fundamental heat transfer behavior with practical design and manufacturing methods becomes increasingly important.

Producing cHEXs using conventional manufacturing (CM) poses significant challenges. CM methods, such as subtractive techniques and forming processes, often require the assembly of components through welding or brazing, making it difficult to achieve leak-proof designs. Additionally, CM's limitations in machining complex geometries further complicate the production of compact, high-performance heat exchangers~\cite{Kaur2021}.

Additive manufacturing (AM) has emerged as a transformative technology for cHEX production, offering unparalleled freedom in design complexity and the ability to manufacture components in a single run~\cite{Jafari2018, Chen2021, Careri2023}. This capability reduces the risk of fluid leakage, minimizes material waste, and supports strength-critical designs~\cite{Kaur2021, Fuchs2022}. Several studies have demonstrated the feasibility of AM for producing compact and efficient heat exchangers~\cite{Saltzman2018}, highlighting its ability to fabricate geometries unattainable with CM~\cite{DaSilva2021,Moon2021, Rao2025, Choudhary2025}. Recent investigations have also employed topology optimization to improve heat transfer in additively manufactured or miniaturized thermal systems~\cite{Marshall2022,  Alshayji2025}, underscoring the convergence of heat-transfer fundamentals and manufacturing innovation. Such approaches show how advanced design tools and AM can be combined to translate thermal optimization concepts into practical engineering solutions.

However, despite these advantages, axial heat conduction remains a critical issue in cHEX design. As heat exchanger volume decreases, axial conduction within the solid wall separating the fluids becomes more significant, reducing thermal effectiveness~\cite{Kroeger1967, Nagasaki2003, Maranzana2004, Fuchs2022} and potentially offsetting the benefits of AM-enabled design enhancements. A better understanding of this effect requires considering both the underlying heat transfer mechanisms and the limitations imposed by AM processes.

While extensive research exists on AM cHEXs and on axial conduction effects, these topics are often addressed independently. Studies on AM processes frequently prioritize high-conductivity materials, such as aluminum and copper~\cite{Careri2023,Hein2021}. Conversely, research on axial wall conduction emphasizes that high thermal conductivity may negatively impact the thermal efficiency of cHEXs~\cite{Bier1993, Stief1999, Borjigin2018}. Additionally, thin plates are known to mitigate axial conduction~\cite{Mitra2023}, but AM processes impose constraints on minimum achievable plate thickness due to limitations in manufacturing resolution, that is, the smallest reliably producible wall thickness specific to each AM technique and material in order to guarantee the cHEX to be leakage-tight. A comprehensive evaluation of axial conduction in AM cHEXs, accounting for these practical limitations, remains unexplored.

%%%%%%%%%%%%%%%%% The deterioration effect of axial conduction on heat exchanger performance/The need for an optimization %%%%%%%%%%%%%%%%%
For general (non-AM) cHEXs, axial conduction has been extensively studied theoretically~\cite{Kroeger1967, Narayanan1999, Aminuddin2017, Raju2017}, numerically~\cite{Ranganayakulu1997, Nellis2003, Quintero2017, Hasan2014}, and experimentally~\cite{Gupta2000, Yang2014}. Researchers have examined its effects on thermal performance under various conditions, emphasizing the importance of parameters such as thermal conductivity, plate thickness, and plate spacing~\cite{Mitra2023, Vera2018}. Optimization studies, such as those by Rogiers and Baelmans~\cite{Rogiers2010} and Buckinx et al.~\cite{Buckinx2013}, have further explored the effect of geometric parameters on the power density. These studies provide valuable insights but often assume ideal manufacturing conditions without considering AM-specific constraints.

%%%%%%%%%%%%%%%%% Our contribution %%%%%%%%%%%%%%%%%
In this paper, we investigate how material thermal properties and AM constraints influence the achievable compactness of counterflow plate heat exchangers. The study focuses on maximizing thermal power density under fixed pressure drop and effectiveness, while accounting for axial wall conduction effects inherent to compact designs. A simplified model of a parallel flat plate heat exchanger is used, with plate length and spacing as design variables and plate thickness treated as a constraint. The impact of AM-specific limitations on wall thickness and fouling risk is systematically evaluated across six different materials. Through this systematic study, we aim to contribute to the understanding of how material properties and their AM-specific limitations affect the compactness of cHEXs. It is shown that the AM-specific manufacturing limits can jeopardize the exploitation of the design freedom offered by this manufacturing technique for cHEXs. 

%%%%%%%%%%%%%%%%% Text structure %%%%%%%%%%%%%%%%%
The text is organized as follows: Section~\ref{sec:methodology} describes the model employed for assessing thermal performance. Section~\ref{sec:results1} discusses unconstrained optimizations for minimizing size across different materials. Section~\ref{sec:results2} revisits these optimizations, incorporating AM and fouling constraints. Section~\ref{sec:results3} evaluates the effect of manufacturing resolution on optimal outcomes. Finally, Section~\ref{sec:discussion} presents a discussion of the results.

%%%
\section{Thermal model} \label{sec:methodology}
The goal of this study is to minimize the size of a parallel-plate counterflow heat exchanger without compromising pressure drop and thermal effectiveness. For a given thermal power, this implies maximizing thermal power density. This section provides a comprehensive description of the analytical thermal model that is used to evaluate the power density, while accounting for axial heat conduction.

\begin{figure}[ht]
    \centering
    \includegraphics[scale=0.35]{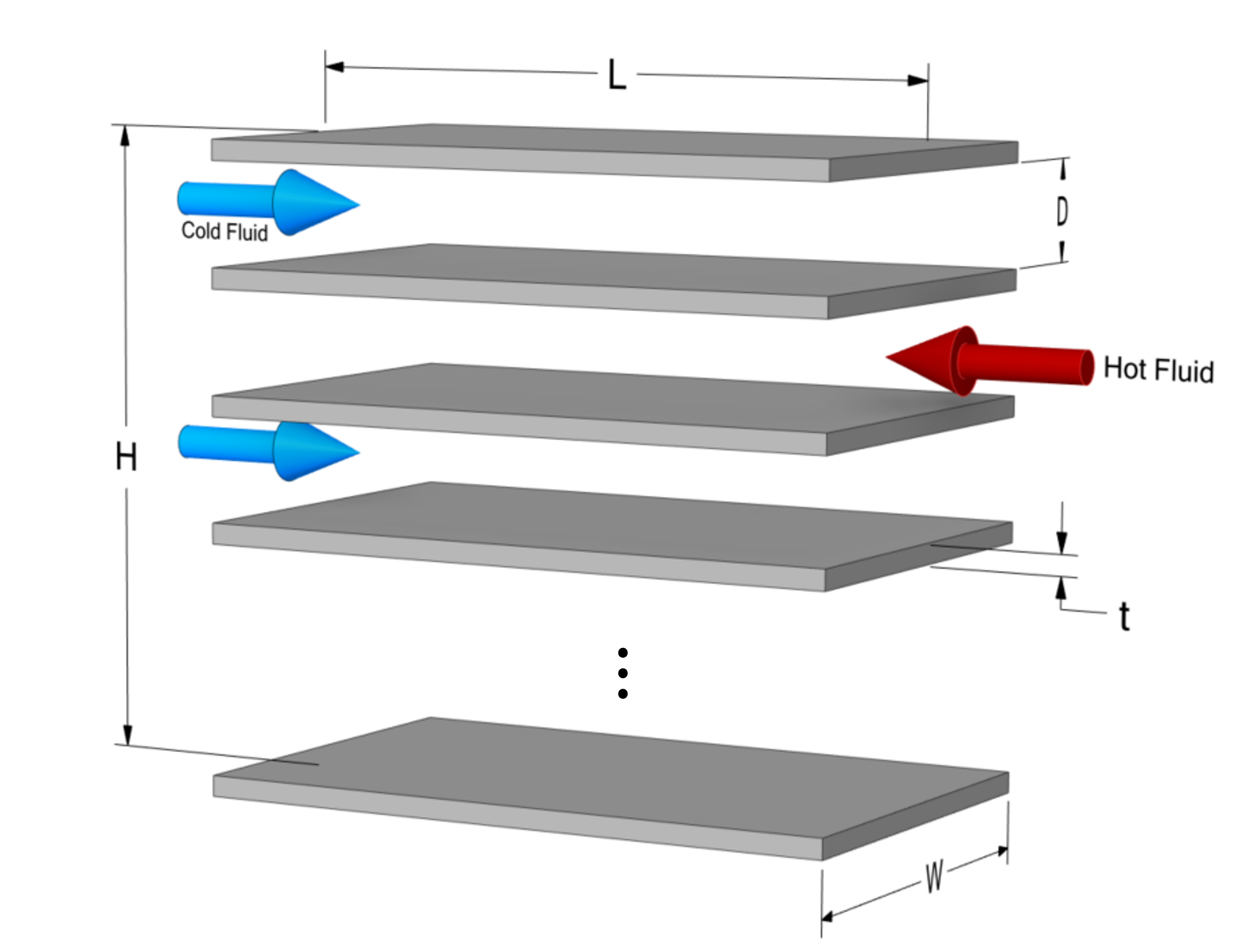}
    \vspace{-4mm}
    \caption{Counterflow parallel flat plate heat exchanger geometry with \(n\) hot and \(n\) cold side channels}
    \label{PFPHE_geo}
\end{figure}

The present investigation is conducted on a simple parallel flat plate heat exchanger (PFPHE), as illustrated in Figure~\ref{PFPHE_geo}. The configuration comprises \(2n+1\) flat, thin plates arranged in parallel, with each plate stacked on top of the preceding one. The plates have a thickness of \(t\) and form \(2n\) fluid channels in a repeating sequence, alternating between cold and hot flows. The channels are of identical spacing \(D\), length \(L\), and width \(W\). Furthermore, this study assumes a perfectly balanced heat exchanger with equal mass flow rates, identical pressure drops, and uniform fluid properties on both the hot and cold sides.

In heat exchangers, power density $\mathcal{D}$ refers to the heat transfer rate per unit volume:
\begin{equation}
    \label{PD}
    \mathcal{D} \equiv \frac{\dot{q}}{H L W}=\frac{\varepsilon \dot{m} c_p \Delta T}{2 n D(1 + t/D) L W},
\end{equation}
where $\varepsilon$ is the effectiveness, representing the ratio of the actual heat transfer rate to the maximum possible heat transfer rate~\cite{Cengel2012}, $\dot{m}$ is the mass flow rate, $c_p$ is the fluid specific heat at constant pressure, and $\Delta T$ is the temperature difference between the hot and cold side inlets. Assuming a sufficiently large number of plates, the total height of the heat exchanger is approximated as \(H=(2n)D+(2n+1)t\approx2 n D(1 + t/D)\).

The mass flow rate $\dot{m}$ at each side is expressed in terms of plate spacing \(D\):
\begin{equation}
    \label{mdot}
    \dot{m} = n \rho V D W,
\end{equation}
where $\rho$ is the density, and $V$ is the bulk velocity. The pressure drop \(\Delta P\) relates to the bulk channel velocity \(V\) through the equation
\begin{equation}
    \label{bulk_velocity}
    V = \frac{2}{f\Re \mu} \frac{D^{2} \Delta P}{L},
\end{equation}
where $f$ is the Fanning friction factor, $\Re$ is the Reynolds number based on hydraulic diameter $2D$, and $\mu$ is the dynamic viscosity.
Upon substituting Eq.~\ref{bulk_velocity} into Eq~\ref{mdot}, the mass flow rate \(\dot{m}\) becomes
\begin{equation}
    \label{mdot2} 
    \dot{m} = \frac{2n\rho}{f\Re\mu} \frac{D^{3} \Delta P}{L} W.
\end{equation}
By substituting Eq.~\ref{mdot2} into Eq.~\ref{PD} and applying algebraic manipulation followed by nondimensionalization, we derive an analytical expression for the nondimensional power density \( Q^{\diamond} \):
\begin{equation}
    \label{NPD} 
    Q^{\diamond} \equiv \frac{\mathcal{D} \mu}{c_p \Delta T \rho \Delta P_{\text{ref}}} = \psi \frac{1}{f\Re} \frac{1}{1 + t^\ast / D^\ast} \varepsilon \left( \frac{D^\ast}{L^\ast} \right)^2.
\end{equation}
Here, the nondimensional geometrical parameters are defined as
\[
L^\ast = \frac{L}{t_{\text{ref}}}, \quad D^\ast = \frac{D}{t_{\text{ref}}}, \quad t^\ast = \frac{t}{t_{\text{ref}}},
\]
where \( t_{\text{ref}} \) is the reference plate thickness, and the factor \( \psi \) is defined as the ratio of actual pressure drop \( \Delta P \) to the reference pressure drop, i.e., \( \psi = \Delta P / \Delta P_{\text{ref}} \). In this work, the reference values \(t_{\text{ref}}\) and \(\Delta P_{\text{ref}}\) correspond respectively to the plate thickness and pressure drop of the initial heat exchanger design, which is described in Table~\ref{use_case}. 

Kroeger~\cite{Kroeger1967} derived an expression to calculate the effectiveness of a balanced counterflow heat exchanger, taking axial heat conduction in the wall into account:
\begin{equation}
    \label{effectiveness}
    \varepsilon = 1 - \frac{1}{1 + \NTU(1 + \M \cdot \phi)/(1 + \M \cdot \NTU)},
\end{equation}
with
\begin{equation}
    \label{phi}
    \phi = \sqrt{\frac{\M \cdot \NTU}{1 + \M \cdot \NTU}} \tanh{\frac{\NTU}{\sqrt{\M \cdot \NTU / (1 + \M \cdot \NTU)}}}.
\end{equation}
The number of transfer units $\NTU$ is a dimensionless parameter that combines the dimensional terms of the overall heat transfer coefficient \(U\), heat transfer surface area \(A\), mass flow rate \(\dot{m}\), and specific heat \(c_p\)~\cite{Zhang2013}, i.e.~\( \NTU = \frac{UA}{\dot{m} c_p} \). Meanwhile, the axial conduction parameter~\( \M = \frac{UA_{ax}}{\dot{m} c_p} \) addresses the effect of heat conduction along the heat exchanger wall~\cite{Kroeger1967, Ranganayakulu1997}, with $A_{ax}$ is the cross-sectional area for axial heat conduction.

For the current PFPHE, the combination of hot and cold side convection and the lateral conduction gives $UA$ as:
\begin{equation}
    \label{UA}
    UA = 2nLW \left( \frac{2}{h} + \frac{t}{k_w}\right)^{-1},
\end{equation}
where $h$ is the convective heat transfer coefficient, and $k_w$ is the thermal conductivity of the plate. The first term within the brackets accounts for the combined convective resistances of the hot and cold flow sides, while the second term addresses the lateral conduction resistance through the plate. By substituting the expression for the convective heat transfer coefficient $h= \frac{\Nu k}{2D}$ with $\Nu$ the Nusselt number and $k$ the fluid thermal conductivity, $UA$ is transformed into
\begin{equation}
    \label{UA2}
    UA = 2nk \frac{L}{D} \frac{\Nu}{4} W \left( 1 + \frac{\Nu}{4} \frac{k}{k_w} \frac{t}{D} \right)^{-1}.
\end{equation}
The axial plate conductance per unit length $UA_{ax}$, which is used in the calculation of the axial conduction parameter \( \M \), is expressed for the present geometry as
\begin{equation}
    \label{UA_ax}
    UA_{ax} = 2n \frac{k_w t}{L}W.
\end{equation}
By substituting Eqs.~\ref{mdot2} and~\ref{UA2} into the definition of~\( \NTU \), an analytical expression for the number of transfer units is obtained:
\begin{equation}
    \label{NTU}
    \NTU = \frac{f\Re \Nu}{4} \left( 1 + \frac{\Nu}{4} \frac{k}{k_w} \frac{t}{D} \right)^{-1} \frac{k}{\rho c_p} \frac{\mu}{\Delta P} \frac{L^2}{D^4}.
\end{equation}
Likewise, by substituting Eqs.~\ref{mdot2} and~\ref{UA_ax} into the definition of the axial conduction parameter \(\M \), we obtain:
\begin{equation}
    \label{M}
    \M = f\Re \frac{k_w}{k} \frac{k}{\rho c_p} \frac{\mu}{\Delta P} \frac{t}{D^3}.
\end{equation}

Next, we express \(\NTU\) and \(\M\) in terms of the previously defined nondimensional parameters: \( L^\ast = L / t_{\text{ref}}, \; D^\ast = D / t_{\text{ref}}, \; t^\ast = t / t_{\text{ref}} \), and \( \psi = \Delta P / \Delta P_{\text{ref}} \). Additionally, we introduce a new dimensionless scaling group \( \Pi = \frac{\Delta P_{\text{ref}} t_{\text{ref}}^2}{\alpha \mu} \), where \( \alpha = \frac{k}{\rho c_p} \) is the fluid thermal diffusivity. The resulting forms of \(\NTU\) and \(\M\) are:
\begin{equation}
    \label{NTU_nd}
    \NTU = \frac{f\Re \Nu}{4} \left( 1 + \frac{\Nu}{4} \frac{k}{k_w} \frac{t^\ast}{D^\ast} \right)^{-1} \frac{1}{\psi \Pi} \frac{{L^\ast}^2}{{D^\ast}^4},
\end{equation}

\begin{equation}
    \label{M_nd}
    \M = f\Re \frac{k_w}{k} \frac{1}{\psi \Pi} \frac{t^\ast}{{D^\ast}^3}.
\end{equation}
These expressions allow for direct evaluation of the thermal effectiveness (via Eq.~\ref{effectiveness}) and power density (via Eq.~\ref{NPD}) as functions of the nondimensional geometrical and operating parameters, capturing the coupled effects of fluid flow, heat transfer, and axial conduction in a compact form. Together, they constitute an analytical model $Q^{\diamond} \left(D^\ast, L^\ast, t^\ast\right)$, as a function of the geometrical parameters $D^\ast$, $L^\ast$, and $t^\ast$.

This study considers heat exchangers operating in the laminar flow regime, which is characteristic of compact, low-Reynolds-number systems. The flow within each channel is assumed to be fully developed and two-dimensional (2D), confined between flat parallel plates. This fully developed flow assumption enables the use of constant values for the friction factor and Nusselt number, and is appropriate for channels with sufficiently long flow lengths relative to the hydraulic diameter. The 2D approximation is justified by the high width-to-spacing ratio that characterizes the geometry of the configuration studied here, which suppresses transverse velocity gradients and minimizes three-dimensional effects.

Under these conditions, the product of the Fanning friction factor and Reynolds number is constant, with \( f\Re = 24 \) for laminar flow between infinite parallel plates~\cite{Rohsenow1998, Mikielewicz2014, Lin2020}. Similarly, the Nusselt number is treated as a known constant. For a fully developed 2D planar channel flow subjected to constant surface heat flux on both walls, \( \Nu = 8.235 \)~\cite{Rohsenow1998, Smith2014}. These values are used as fixed inputs in the present model and reflect typical conditions for idealized laminar channel flows. Additional 3D conjugate heat transfer (CHT) simulations have been conducted using ANSYS Fluent~\cite{Ansys2024fluentr1guide} to compare the analytical model predictions with high-fidelity numerical results. These results, provided in the supplementary material, show close agreement with the analytical predictions, with less than 0.52\% relative deviation in thermal effectiveness (see Figure~S1 for boundary conditions and Table~S1 for simulation results in the supplementary material). This comparison supports the validity of the modeling assumptions adopted in this study. 

\section{Material selection for maximal power density} \label{sec:results1}

A wide variety of materials prove suitable for the manufacturing of heat exchangers, and the choice is typically driven by the requirements and operating conditions of the device. Stainless steel, for example, finds widespread applications in heat exchanger manufacturing due to its resistance to corrosion and high temperatures~\cite{ZadiMaad2018}. Copper is often preferred in heat transfer applications due to its superior thermal conductivity~\cite{Careri2023}. Aluminum stands out as a common choice due to its lightweight, excellent thermal conductivity, and affordability compared to other metals~\cite{Cantor2015}. Ceramic materials are increasingly considered advantageous for heat exchangers due to their temperature and corrosion resistance, outperforming traditional metallic variants~\cite{Sommers2010}. Beyond metals and ceramics, polymers also offer advantages in heat exchanger manufacturing, such as cost-effectiveness, lightweight construction, and resistance to corrosion and fouling~\cite{Hein2021}.
Considering this diversity, the current study investigates the effect of six different materials on the maximum power density. These materials are copper (with a thermal conductivity, $k_{\text{Cu}} = 398~\text{W/m} \cdot \text{K}$), aluminum ($k_{\text{Al}} = 237~\text{W/m} \cdot \text{K}$), aluminum nitride (AlN, $k_{\text{AlN}} = 180~\text{W/m} \cdot \text{K}$), aluminum oxide (Al\textsubscript{2}O\textsubscript{3}, $k_{\text{Al\textsubscript{2}O\textsubscript{3}}} = 27~\text{W/m} \cdot \text{K}$), austenitic steel ($k_{\text{A.st}} = 20~\text{W/m} \cdot \text{K}$), and a typical plastic ($k_{\text{Pl}} = 0.2~\text{W/m} \cdot \text{K}$). 

First, we examine the effect of different materials on the maximum power density at a given effectiveness, without accounting for manufacturing and fouling constraints. To achieve this, we formulate an optimization problem that maximizes power density for each material, characterized by its thermal conductivity \(k_w\), by varying plate spacing and channel length. To avoid unrealistic solutions where the optimizer reduces the wall thickness to zero, we impose a geometric constraint that links wall thickness in proportion to plate spacing: \( t = \gamma D \), where \( \gamma \) is a fixed constant. This prevents physically degenerate designs by ensuring the wall always retains finite thickness. The optimization problem is then written in dimensionless form as
\begin{equation}
    \label{maxQ}
    \begin{aligned}
    \max_{D^\ast > 0,\ L^\ast > 0,\ t^\ast > 0} & \qquad Q^{\diamond} \left(D^\ast, L^\ast, t^\ast \right), \\
    \text{s.t.} & \qquad \varepsilon\left(D^\ast, L^\ast, t^\ast; k_w/k\right) = \varepsilon_d, \\
     & \qquad t^\ast = \gamma \cdot D^\ast, \\
    \end{aligned}
\end{equation}
where \(\varepsilon_d\) is the design effectiveness. 
This formulation follows the same optimization structure introduced by Rogiers and Baelmans~\cite{Rogiers2010}, who also maximized power density under a fixed effectiveness constraint. The nondimensional parameters used here differ, primarily due to the explicit treatment of wall thickness, which enables the integration of additional constraints in future design stages.

The nonlinear optimization problem is solved numerically for each of the six materials using Python’s \texttt{SciPy} optimization library, specifically the \texttt{minimize} function, which supports various constrained-solvers~\cite{Virtanen2020}. In this study, the Sequential Least Squares Programming (SLSQP) algorithm is employed to compute the optimal values of \( D^\ast \) and \( L^\ast \), while satisfying the imposed constraints. This numerical procedure is used consistently for all optimization problems discussed in the paper. To ensure comparability with the baseline heat exchanger (Table~\ref{use_case}), the design effectiveness is fixed at \( \varepsilon_d = 0.79 \), and the proportionality constant is set to \( \gamma = 0.16 \), corresponding to the original \( t^\ast / D^\ast \) ratio. The thermophysical properties of the fluid, listed in Table~\ref{use_case}, are evaluated at the mean of the inlet temperatures of the hot and cold fluid and are assumed constant throughout the analysis. The pressure drop is matched to the baseline to ensure a fair comparison across all cases.
\begin{table}[!ht]
    \centering
    \caption{Working fluid and initial heat exchanger properties. The plate thickness and pressure drop listed here correspond to the reference values $t_{\text{ref}}$ and $\Delta P_{\text{ref}}$ used throughout the text.}
    \begin{threeparttable}
        % \small
        \begin{tabular}{@{}l l@{}}
            \toprule
            Property & Value \\
            \midrule
            Working fluid                             & Air \\
            Fluid density, \(\rho\)                   & \SI{1.060}{\kilo\gram\per\meter\cubed} \\
            Fluid isobaric specific heat, \(C_p\)     & \SI{1.008}{\kilo\joule\per\kilo\gram\per\kelvin} \\
            Fluid dynamic viscosity, \(\mu\)          & \SI{19.99e-6}{\pascal\second}\\            
            Fluid thermal conductivity, \(k\)         & \SI{28.8}{\milli\watt\per\meter\per\kelvin}\\
            Cold fluid inlet temperature, \(T_{in,c}\)& \SI{20}{\degreeCelsius} \\
            Hot fluid inlet temperature, \(T_{in,h}\) & \SI{100}{\degreeCelsius}\\
            Pressure drop, \(\Delta P\)               & \SI{170}{\pascal}\\ 
            Effectiveness, \(\varepsilon\)            & 0.79 \\  
            Length, \(L\) & \SI{158.0}{\milli\meter} \\
            Width, \(W\) & \SI{95.0}{\milli\meter} \\
            Plate spacing, \(D\) & \SI{1.0}{\milli\meter} \\
            Plate thickness, \(t\) & \SI{0.16}{\milli\meter} \\
            Material                                  & Austenitic steel \\            
            \bottomrule
        \end{tabular}
    \end{threeparttable}
    \label{use_case}
\end{table}
The optimization results are summarized in Table~\ref{opt_without_manuf_const}, which reports the nondimensional geometrical parameters \( L^\ast \), \( D^\ast \), and \( t^\ast \) that maximize power density for each material. The corresponding nondimensional power densities \( Q^{\diamond} \) are also listed, along with the Improvement Factor (IF), defined as the ratio of the optimized power density to that of the baseline design, referred to as ``austenitic steel (initial)".

\begin{table}[!ht]
    \centering
    \caption{Nondimensional properties of initial and optimized heat exchangers without manufacturing constraints, with associated improvement factors. All geometrical values are scaled by the reference wall thickness \( t_{\text{ref}} = \SI{0.16}{\milli\meter} \).}
    \begin{threeparttable}
        % \resizebox{\textwidth}{!}{
            \begin{tabular}{lS[table-format=3.2]rS[table-format=3.2]rS[table-format=1.2]rS[table-format=1.4]rS[table-format=4.2]rS[table-format=6.1]}
                \toprule
                {\textbf{Material}} & {\textbf{$k_w/k$}} & {\textbf{$L^\ast$}} & {\textbf{$D^\ast$}} & {\textbf{$t^\ast$}} & {\textbf{\( Q^{\diamond} \times 10^6\)}} & {\textbf{IF}} \\ 
                \midrule
                Austenitic steel (Initial)  & 694.44  & 987.50  & 6.250 & 1.000 & 1.138 & 1.00 \\
                Plastic                     & 6.94  & 0.70  & 0.137 & 0.022 & 1095.615 & 962.44 \\
                Austenitic steel            & 694.44  & 68.44  & 1.375 & 0.220 & 11.470 & 10.08 \\
                Ceramic (Al\textsubscript{2}O\textsubscript{3})             & 937.50  & 91.51  & 1.582 & 0.253 & 8.495 & 7.46 \\
                Ceramic (AlN)               & 6250.00  & 615.53 & 4.123 & 0.660 & 1.275 & 1.12 \\
                Aluminum                    & 8229.17  & 808.22 & 4.718 & 0.755 & 0.968 & 0.85 \\
                Copper                      & 13819.44  & 1361.25 & 6.133 & 0.981 & 0.577 & 0.51 \\          
                \bottomrule
            \end{tabular}
        % }
    \end{threeparttable}
    \label{opt_without_manuf_const}
\end{table}

In the optimal configurations, the nondimensional geometrical parameters \(L^\ast\), \(D^\ast\), and \(t^\ast\) increase with the material's thermal conductivity, leading to a reduction in nondimensional power density \(Q^{\diamond}\). For instance, the plastic heat exchanger, which has the lowest plate conductivity, achieves the highest performance, with an IF of approximately 962 relative to the initial design.  In contrast, materials with high conductivity, such as aluminum and copper, result in lower power densities than the baseline. Copper, in particular, exhibits the lowest performance, with an IF of 0.51, indicating a 49\% reduction compared to the initial design.

To interpret the optimized nondimensional designs in physical terms, each configuration can be translated into dimensional quantities using the wall thickness of the initial heat exchanger configuration as the reference wall thickness \( t_{\text{ref}} \). The nondimensional parameters \( L^\ast \), \( D^\ast \), and \( t^\ast \) are rescaled to obtain the corresponding dimensional values \( L \), \( D \), and \( t \). To maintain consistency with the baseline thermal power, the number of layers \( n \) and the width \( W \) are selected such that the product with the mass flow rate per unit width for a single channel, \( \dot{m}^{\prime}_\text{single} \), satisfies~$(nW\dot{m}^{\prime}_\text{single})_{\text{initial}} = (nW\dot{m}^{\prime}_\text{single})_{\text{optimal}}.$ This approach ensures that the dimensional designs deliver the same thermal power as the baseline configuration. The corresponding dimensional properties, including \( L \), \( D \), \( t \), \( W \), \( n \), \( \dot{m}^{\prime}_\text{single} \), and \( \mathcal{D} \), are provided in the supplementary material (Table~S2).

Without any constraints on plate thickness and spacing, the optimal configurations feature \( t^\ast \) values ranging from 0.022 to 0.981. These correspond to dimensional thicknesses as low as \SI{3.5}{\micro\meter} and up to \SI{157}{\micro\meter}, based on the reference wall thickness \( t_{\text{ref}} = \SI{0.16}{\milli\meter} \). Such thin structures raise concerns about structural integrity and manufacturability. The AM processes cannot reliably ensure the producibility or reproducibility of features at these small scales, and the accuracy error increases as the component thickness decreases~\cite{Careri2023}. This challenge is especially pronounced for copper and its alloys due to their high surface reflectivity and low laser absorption~\cite{Campagnoli2021, Tiberto2019, Jiang2021}. Additionally, \( D^\ast \) can be as low as 0.137 in the optimal heat exchangers, corresponding to a dimensional plate spacing of \SI{22}{\micro\meter}, which increases the risk of fouling. These limitations motivate a revised optimization problem that incorporates manufacturability constraints and accounts for practical design considerations such as fouling risk.

\section{Design optimization considering additive manufacturing and fouling constraints}\label{sec:results2}

In cHEX design, thinner plates are generally preferred because they reduce axial heat conduction and minimize lateral conduction resistance, enhancing thermal performance. However, the minimum achievable plate thickness is limited by the capabilities of the selected manufacturing process and material. The goal of this section is to account for constraints related to manufacturing, fouling, and structural limitations when designing a cHEX for maximal compactness for a given pressure drop and effectiveness. To this end, we will introduce constraints on plate thickness and spacing into the optimal design problem. In the previous section, the nondimensional plate geometry was defined using a fixed ratio between plate thickness and spacing (\( t^\ast/D^\ast \)), which implicitly restricted the ability to impose a specific, manufacturable design plate thickness \( t_d \). To explicitly introduce the minimal plate thickness, the original optimization problem in Eq.~\ref{maxQ} is reformulated  as follows:
\begin{equation}
    \label{maxQ_t_cons} 
    \begin{aligned}
    \max_{D^\ast > 0,\ L^\ast > 0,\ t^\ast > 0} & \qquad Q^{\diamond} \left(D^\ast, L^\ast, t^\ast \right), \\
    \text{s.t.} & \qquad \varepsilon\left(D^\ast, L^\ast, t^\ast; k_w/k\right) = \varepsilon_d, \\
     & \qquad t^\ast = t_{d}/t_{\text{ref}}, \\
     & \qquad D^\ast \geq D_{\text{min}}/t_{\text{ref}}. \\
    \end{aligned}
\end{equation}
In addition to the plate thickness, a restriction on the plate spacing is imposed to reduce the risk of fouling. Specifically, the plate spacing \( D \) must remain greater than a prescribed minimum value \( D_{\text{min}} \), based on practical operating considerations. In the nondimensional formulation, this is enforced through the constraint \( D^\ast \geq D_{\min} / t_{\text{ref}} \).

\subsection{Thickness-constrained power density optimization for a design effectiveness of 0.79}
\begin{figure}[ht]
    \centering
    \includegraphics[scale=0.65]{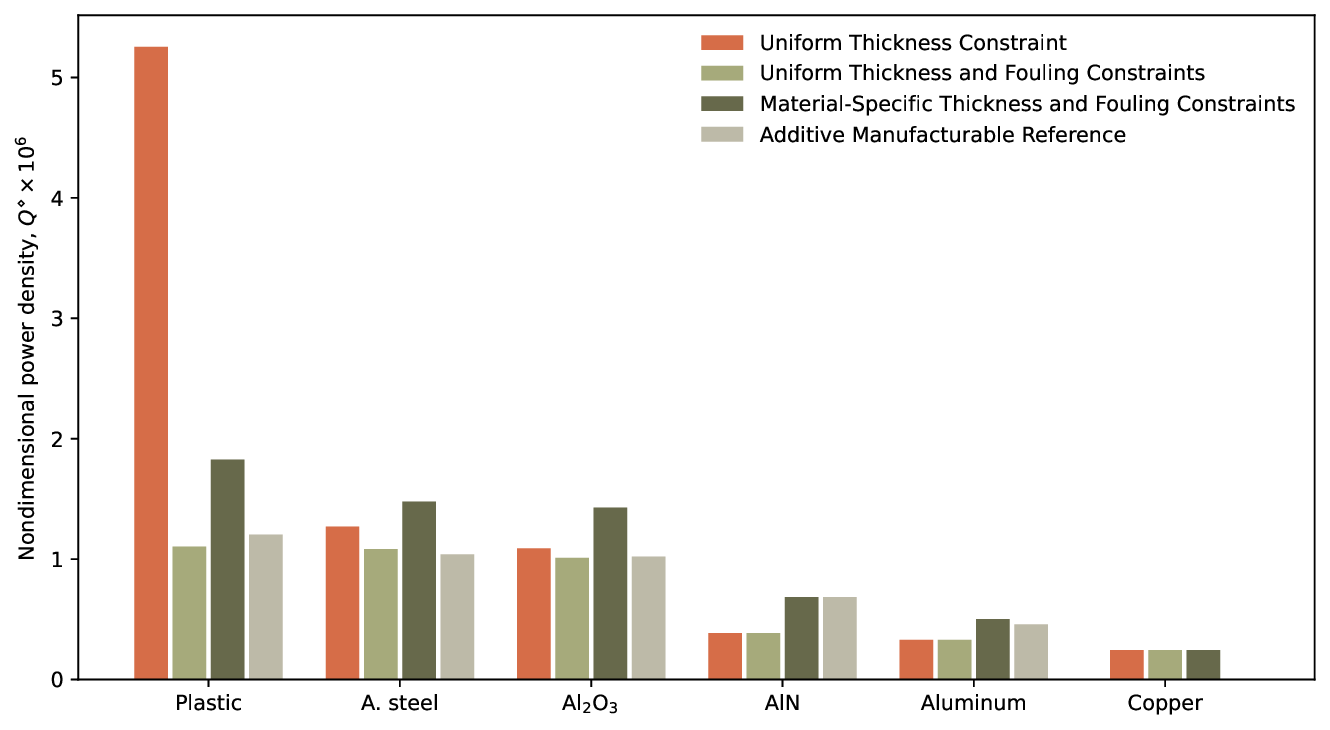}
    \vspace{-4mm}
    \caption{Maximum nondimensional power densities obtained from optimizations with~\(\varepsilon_d~=~0.79\)}
    \label{opt_results}
\end{figure}

The optimizations in this section use the same reference values, fluid properties, pressure drop, and effectiveness as the baseline heat exchanger, outlined in Section~\ref{sec:results1} and detailed in Table~\ref{use_case}. The results of these optimizations are illustrated in Figure~\ref{opt_results}, which compares three distinct optimization strategies aimed at maximizing power density. 

The first strategy, labeled ``\textit{Uniform Thickness Constraint}", applies a uniform \(t_d\)=0.5 mm across all materials, with no constraint on plate spacing. The second strategy, labeled ``\textit{Uniform Thickness and Fouling Constraints}", introduces a fouling constraint that imposes $D_{min}$=0.8 mm, while still applying the same uniform plate thickness of 0.5 mm across all materials. The third strategy, labeled ``\textit{Material-Specific Thickness and Fouling Constraints}", also incorporates both plate thickness and fouling constraints simultaneously. However, this approach applies a material-specific thickness constraint to account for variations in manufacturing resolution in AM between different materials and processes. 

To address AM resolutions for ceramic and metal heat exchangers, the plate thickness is determined based on the melt pool size, a characteristic of the manufacturing process. It is assumed that a sealed plate requires a thickness of twice the melt pool size. Consequently, the following plate thickness values are used: 0.5 mm for copper~\cite{Jadhav2021}, 0.3 mm for aluminum~\cite{Ghasemi2021}, and 0.25 mm for aluminum nitride (AlN), aluminum oxide (Al\textsubscript{2}O\textsubscript{3})~\cite{Yao2023}, and austenitic steel~\cite{Yang2021}. These values are characteristic of the Laser Powder Bed Fusion (LPBF) process, which is the most common technique for AM of metals and ceramics. 
For plastic heat exchangers, the plate thickness is not derived from melt pool size, as LPBF is not typically used for plastic printing. Instead, Material Extrusion (MEX), Vat Photopolymerization (VPP), and Sheet Lamination (SHL) processes are widely used~\cite{Bhatia2023, Ligon2017}. While these techniques can achieve thin walls with resolutions below 0.1 mm, a plate thickness of 0.1 mm is assumed in this study. This value was selected to balance manufacturing feasibility with structural integrity, as heat exchangers with wall thicknesses below 0.1 mm have been shown to be prone to plastic deformation~\cite{Rua2015}. 

In addition to the three optimization strategies, Figure~\ref{opt_results} also presents nondimensional power density values for reference heat exchanger designs that are additive manufacturable. These are labeled as ``\textit{Additive Manufacturable Reference}'' in the figure. In these designs, the plate spacing of the baseline heat exchanger is preserved, while the plate thickness is adjusted to reflect material-specific manufacturability constraints. To ensure comparability across materials, the channel length is modified as necessary so that each design achieves the same pressure drop of 170\,Pa and the target effectiveness of 0.79. This definition aligns with the classical design methodology, where performance targets are met by varying channel length while maintaining the plate spacing~\cite{Kays1984}.  
However, for copper, even with length adjustments, the design cannot attain the target effectiveness. The reason is that its asymptotic high-NTU effectiveness is limited to \( \varepsilon_{\mathrm{NTU}\rightarrow\infty} = \frac{\M+1}{2\M+1} \approx 0.74,\) with \(\M \approx 0.53\) obtained from Eq.~\ref{M}, which is below the desired effectiveness of 0.79. Therefore, no Additive Manufacturable Reference design in copper is included in the figure.
Consequently, the Additive Manufacturable Reference designs provide a consistent baseline for evaluating the effect of material properties on power density, while highlighting limitations in certain cases.
 
To provide additional context for the designs presented in Figure~\ref{opt_results}, Table~\ref{opt_with_manuf_const} lists the nondimensional geometrical parameters and corresponding nondimensional power densities for each optimization strategy and the additive manufacturable reference designs. The geometrical values include nondimensional plate thickness \(t^\ast\), plate spacing \(D^\ast\), and heat exchanger length \(L^\ast\), defined relative to the reference wall thickness \(t_{\text{ref}} = \SI{0.16}{\milli\meter}\). The dimensional equivalents are provided in Table~S3 in the supplementary material.

\begin{table}[!ht]
    \centering
    \caption{Nondimensional geometrical parameters and power densities for different optimization strategies and reference cases with \( \varepsilon_d = 0.79 \). All geometrical values are scaled by the reference plate thickness \( t_{\text{ref}} = \SI{0.16}{\milli\meter} \).}
    \begin{threeparttable}
        \small
        \begin{tabular}{>{\raggedright\small}m{0.34\linewidth}>{\raggedright\small}m{0.22\linewidth}rrrrr}
            \toprule
            \textbf{Case} & \textbf{Material} & \textbf{\( t^\ast \)} & \textbf{\( D^\ast \)} & \textbf{\( L^\ast \)} & \textbf{\( Q^{\diamond} \times 10^{6} \)} \\ 
            \midrule
            \multirow{6}{*}{\thead{\small Uniform Thickness  \\ \small Constraint}} 
            & Plastic            & 3.125 & 0.946 & 36.10 & 5.254 \\
            & Austenitic steel   & 3.125 & 3.725 & 442.42 & 1.271 \\
            & Ceramic (Al\textsubscript{2}O\textsubscript{3}) & 3.125 & 4.278 & 565.46 & 1.090 \\
            & Ceramic (AlN)      & 3.125 & 7.670 & 1890.17 & 0.386 \\
            & Aluminum           & 3.125 & 8.413 & 2272.68 & 0.329 \\
            & Copper             & 3.125 & 9.961 & 3197.05 & 0.244 \\ 
            \midrule
            \multirow{6}{*}{\thead{\small Uniform Thickness \\ \small and Fouling Constraints}} 
            & Plastic            & 3.125 & 5.00\textsuperscript{*} & 677.65 & 1.104 \\
            & Austenitic steel   & 3.125 & 5.00\textsuperscript{*} & 684.29 & 1.083 \\
            & Ceramic (Al\textsubscript{2}O\textsubscript{3}) & 3.125 & 5.00\textsuperscript{*} & 708.41 & 1.010 \\
            & Ceramic (AlN)      & 3.125 & 7.670 & 1890.17 & 0.386 \\
            & Aluminum           & 3.125 & 8.413 & 2272.68 & 0.329 \\
            & Copper             & 3.125 & 9.961 & 3197.05 & 0.244 \\
            \midrule
            \multirow{6}{*}{\thead{\small Material-Specific Thickness \\ \small and Fouling Constraints}} 
            & Plastic            & 0.625 & 5.00\textsuperscript{*} & 633.37 & 1.826 \\
            & Austenitic steel   & 1.563 & 5.00\textsuperscript{*} & 652.01 & 1.477 \\
            & Ceramic (Al\textsubscript{2}O\textsubscript{3}) & 1.563 & 5.00\textsuperscript{*} & 662.81 & 1.429 \\
            & Ceramic (AlN)      & 1.563 & 6.024 & 1177.34 & 0.685 \\
            & Aluminum           & 1.875 & 7.175 & 1636.86 & 0.502 \\
            & Copper             & 3.125 & 9.961 & 3197.05 & 0.244 \\
            \midrule
            \multirow{5}{*}{\thead{\small Additive Manufacturable \\ \small Reference}} 
            & Plastic            & 0.625 & 6.250 & 986.01 & 1.204 \\
            & Austenitic steel   & 1.563 & 6.250 & 995.67 & 1.039 \\
            & Ceramic (Al\textsubscript{2}O\textsubscript{3}) & 1.563 & 6.250 & 1003.94 & 1.022 \\ 
            & Ceramic (AlN)      & 1.563 & 6.250 & 1227.10 & 0.684 \\
            & Aluminum           & 1.875 & 6.250 & 1469.77 & 0.458 \\
            \bottomrule
        \end{tabular}
        \begin{tablenotes}
            \footnotesize
            \item[*] Fouling constraint is active.
        \end{tablenotes}
    \end{threeparttable}
    \label{opt_with_manuf_const}
\end{table}

As shown in Figure~\ref{opt_results}, under the ``\textit{Uniform Thickness Constraint}'' strategy, where the plate thickness is uniformly set to 0.5\,mm (\( t^\ast = 3.125 \)) and the fouling constraint is not considered, plastic, the material with the lowest thermal conductivity, outperforms all other designs. The geometrical dimensions of this design are significantly smaller than those of the others (see Table~\ref{opt_with_manuf_const}). When the fouling constraint is introduced while still keeping the plate thickness uniform at 0.5\,mm for all materials, the performance of plastic remains slightly better than that of austenitic steel and Al\textsubscript{2}O\textsubscript{3}. However, in the ``\textit{Material-Specific Thickness and Fouling Constraints}'' strategy, where material-specific resolutions are explicitly accounted for, plastic again demonstrates a clearer performance advantage in power density due to its ability to utilize thinner plates.

In contrast, despite having the highest thermal conductivity among the materials considered, copper exhibits the lowest power density across all optimization strategies in Figure~\ref{opt_results}. This outcome highlights a broader trend observed in the results: heat exchangers made from high-conductivity materials such as copper, aluminum, and AlN consistently require increased lengths and plate spacings to achieve the target effectiveness of 0.79. As shown in Table~\ref{opt_with_manuf_const}, the fouling constraint remains inactive for these materials in all strategies, since their optimal spacings are naturally above the minimum threshold. However, their plate spacings are slightly smaller when material-specific thicknesses are used, compared to the uniform-thickness cases. In general, designs with thicker plates and high-conductivity materials result in larger heat exchangers, as longer channels and wider gaps are required to offset axial wall conduction and maintain the target effectiveness.

\subsection{Influence of design effectiveness on optimal heat exchanger}
\begin{figure}[ht]
    \centering

    \includegraphics[scale=0.62]{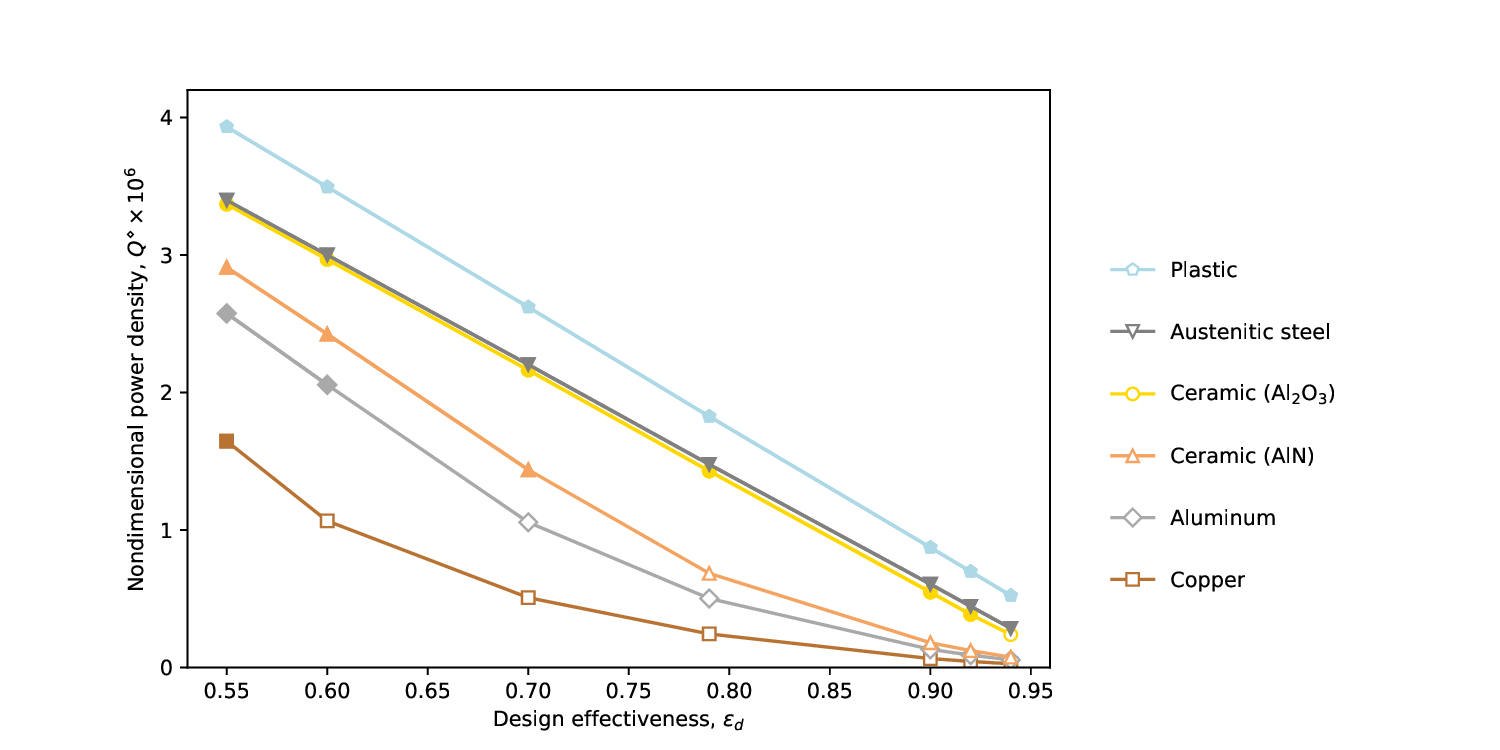}
    \vspace{-4mm}
    \caption{Maximum nondimensional power density with varying effectiveness (fouling constraint active if filled, inactive if open)}
    \label{opt_results_varying_eff}
\end{figure}

In the previous optimizations, a design effectiveness of 0.79 was used to ensure consistency with the baseline heat exchanger (see Table~\ref{use_case}). These studies demonstrated that both material selection and plate thickness significantly influence cHEX power density. To further explore these effects, this section investigates how varying the design effectiveness influences material selection and power density under AM-imposed plate thickness constraints. The ``\textit{Material-Specific Thickness and Fouling Constraints}" optimization strategy is repeated across a range of $\varepsilon_d$, from 0.55 to 0.94, to examine the interplay between effectiveness, material type, and power density. The outcomes of these optimizations for all materials are illustrated in Figure~\ref{opt_results_varying_eff}.

As expected, the nondimensional power density increases as the design effectiveness decreases for all material cases, indicating that lower effectiveness results in smaller heat exchangers. While this increase in power density is relatively linear across the effectiveness range for low thermal conductivity materials, it becomes more pronounced for higher thermal conductivity materials. This suggests that the dimensions of high thermal conductivity heat exchangers are more sensitive to changes in design effectiveness compared to low thermal conductivity heat exchangers. As a result, the fouling constraint, which is inactive for high thermal conductivity heat exchangers at high effectiveness levels, becomes active at low effectiveness levels due to the reduction in plate spacing (see copper, aluminum, and AlN heat exchangers in Figure~\ref{opt_results_varying_eff}).

Figure~\ref{opt_results_varying_eff} further demonstrates that, while plastic consistently achieves the highest power density across the entire range of design effectiveness values considered, nondimensional power density decreases as material thermal conductivity increases. Notably, the optimal material choice remains unaffected by changes in design effectiveness.

%%%%
\section{Impact of the plate thickness on maximal power density}\label{sec:results3}
The previous section determined the plate thickness based on AM limits for metal and ceramic heat exchangers. However, these limits can be significantly exceeded with conventional manufacturing (CM) techniques. For instance, the literature reports microchannels made from stainless steel plates as thin as 0.051 mm~\cite{Mahabunphachai2008}, or even copper plates as thin as 0.003 mm~\cite{Joo2004}, achieved through hydroforming and pressing, respectively. Such thicknesses are far beyond the level of precision currently achievable with AM. This raises the question whether the optimal material choice changes if AM resolution is improved, or whether current AM methods will continue to act as a bottleneck in cHEX design.

\begin{figure}
    \centering
    \begin{subfigure}[b]{0.8\textwidth}
        \centering
        \includegraphics[width=1.0\textwidth]{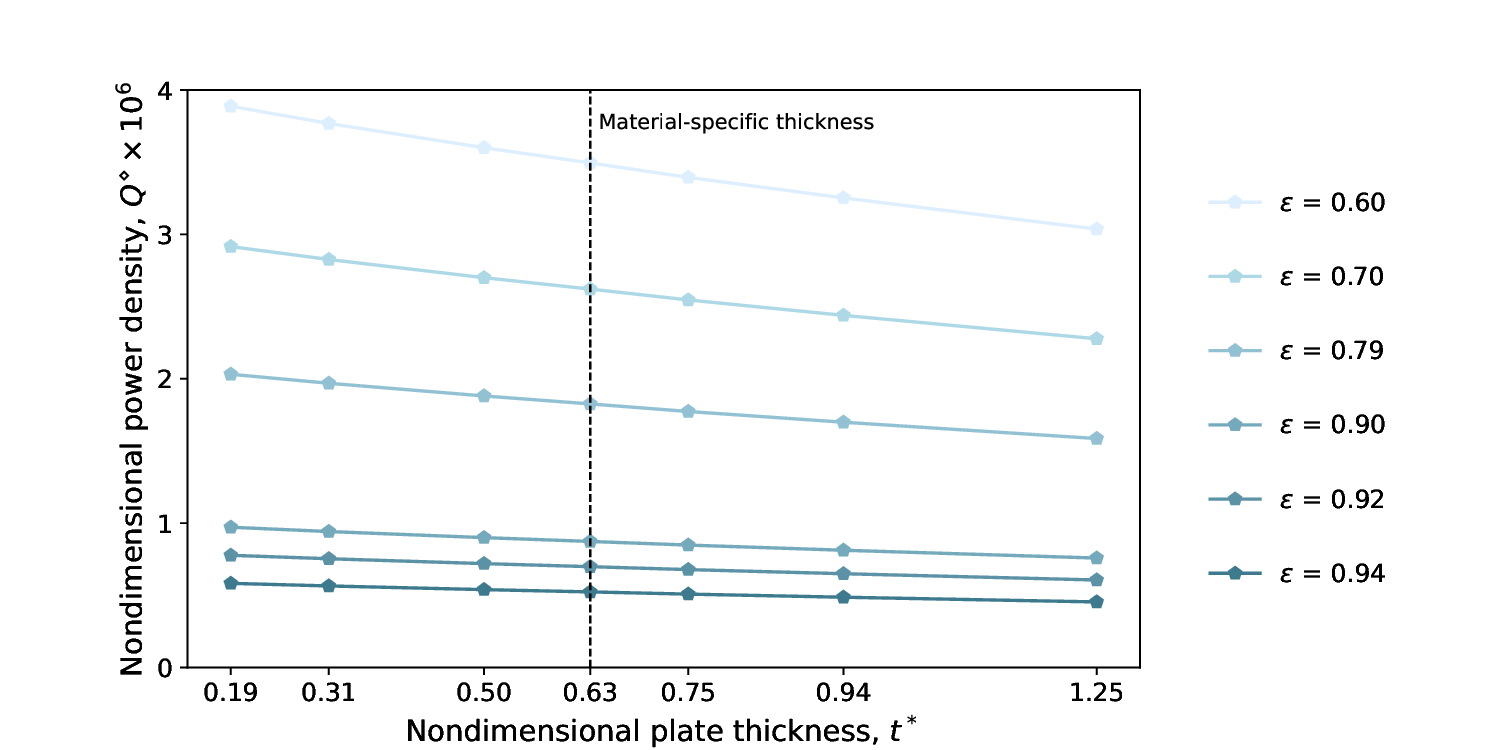}
        \caption{Plastic}
        \label{NPD_vs_t_d_plastic}
    \end{subfigure}
    \vfill
    \begin{subfigure}[b]{0.8\textwidth}
        \centering
        \includegraphics[width=1.0\textwidth]{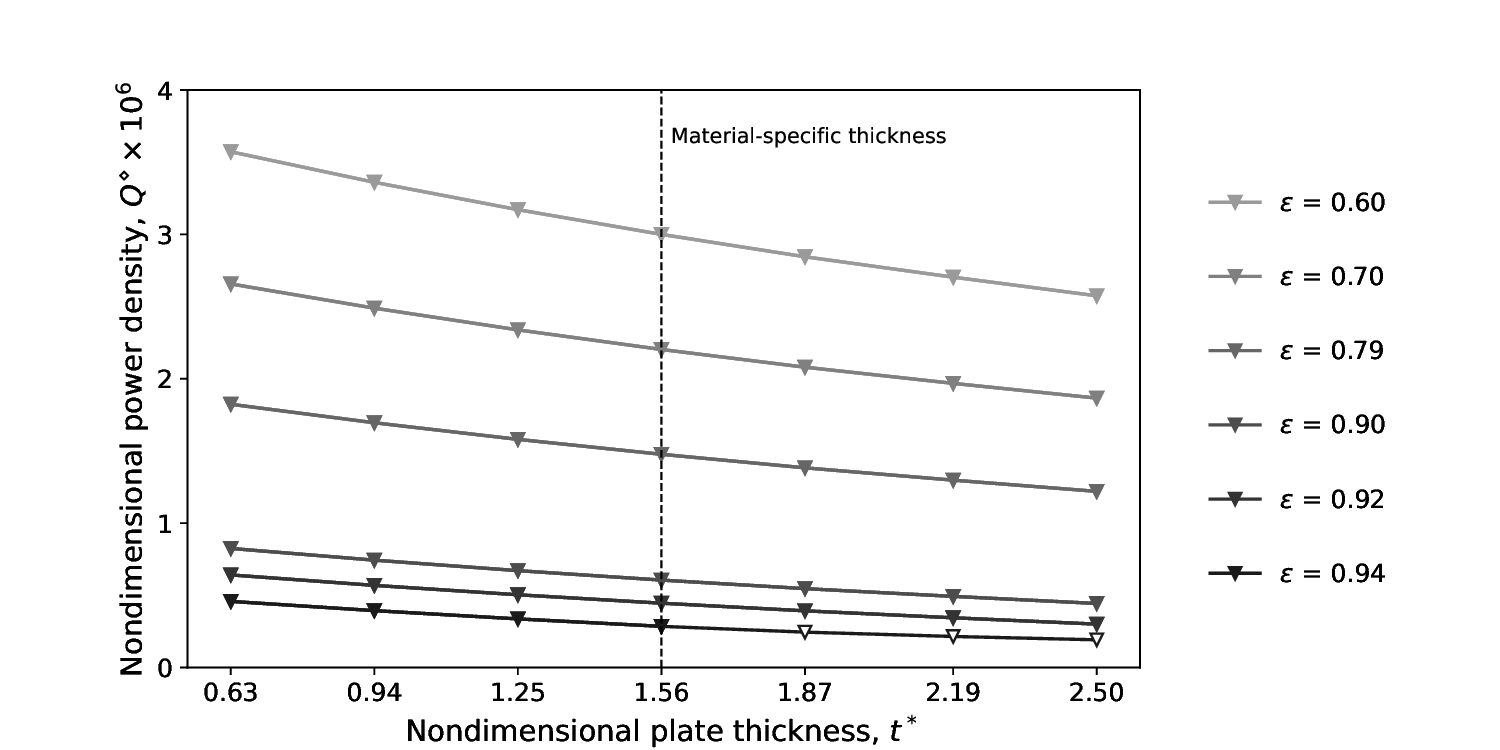}
        \caption{Austenitic steel}
        \label{NPD_vs_t_d_aus_steel}
    \end{subfigure}
    \vfill
    \begin{subfigure}[b]{0.8\textwidth}
        \centering
        \includegraphics[width=1.0\textwidth]{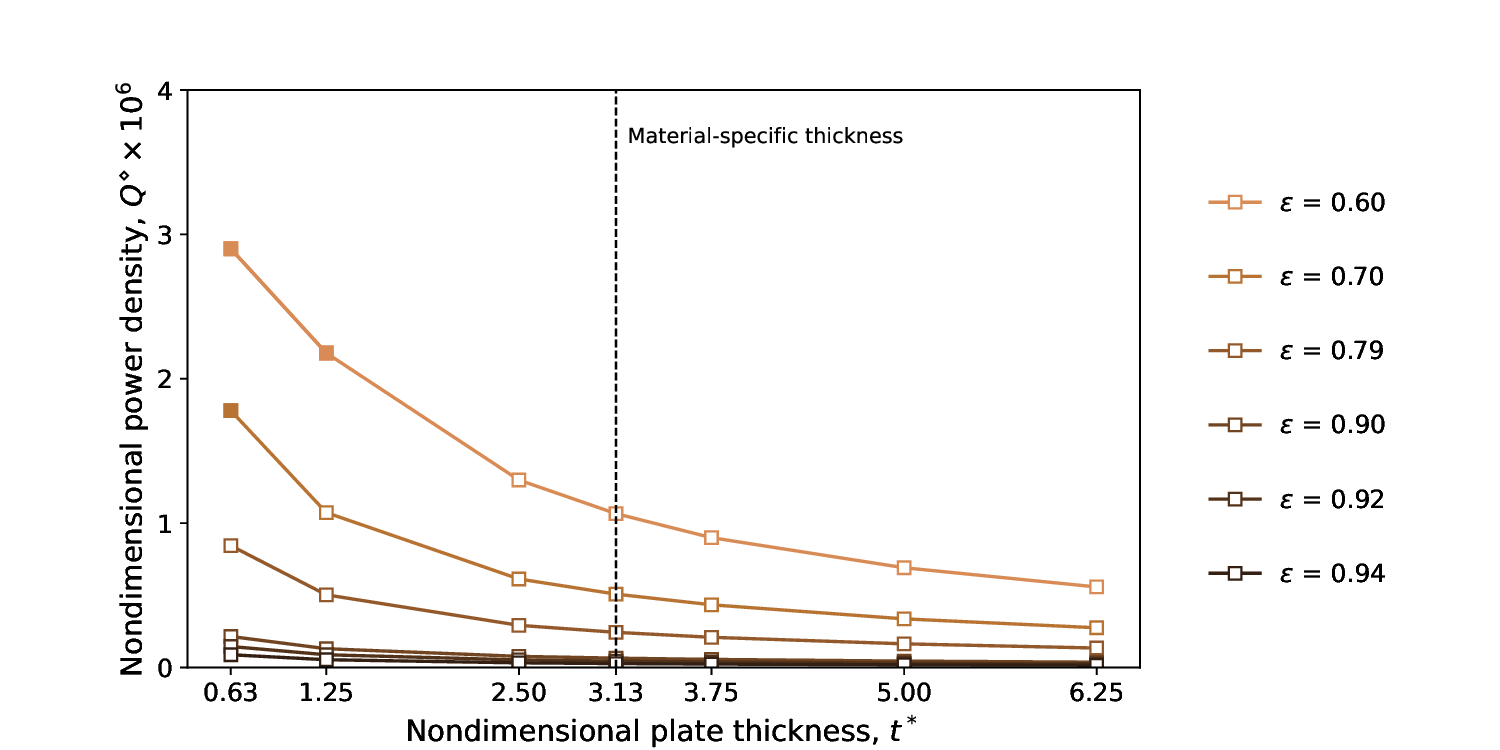}
        \caption{Copper}
        \label{NPD_vs_t_d_copper}
    \end{subfigure}
    \caption{Nondimensional power density variation with nondimensional plate thickness \(t^{\ast}\) 
    (fouling constraint active if filled, inactive if open). The vertical dashed line indicates the 
    material-specific plate thickness used in Section~\ref{sec:results2}.}
    \label{NPD_vs_t_d}
\end{figure}

To address this question, optimization studies are conducted for plastic, austenitic steel, and copper for varying minimal plate thicknesses and design effectiveness levels, while maintaining the fouling constraint fixed. The range of plate thicknesses used in this section is set around the material-specific values from the previous section. For plastic, the material-specific thickness was determined based on structural concerns rather than the limitations of AM. Nonetheless, the same procedure is applied to the plastic heat exchanger to examine the effect of plate thickness on its power density. The results of these optimizations are depicted in Figure~\ref{NPD_vs_t_d}. Each point in this figure corresponds to an optimization run.

Figure~\ref{NPD_vs_t_d} illustrates that the limitations of AM in producing thin features restrict the maximum power density and, consequently, the minimum achievable size of the heat exchanger. For all effectiveness values examined across the three materials, reducing the plate thickness leads to an increase in maximum power density. This increase is relatively modest for plastic and austenitic steel but significantly more noticeable for copper. This suggests that advancements in AM capabilities, particularly in thin-wall manufacturing, would have a greater impact on reducing the volume of copper heat exchangers compared to those made from other materials.

Despite copper’s higher thermal conductivity, it still results in a larger heat exchanger compared to austenitic steel and plastic, as shown in Figure~\ref{NPD_vs_t_d}. When the same plate thickness of $t^\ast=0.63$ ($t = 0.1\,\text{mm}$) or $t^\ast=1.25$ ($t = 0.2\,\text{mm}$) is used in the optimizations for all materials, copper consistently yields the lowest power density across all effectiveness values. In contrast, the austenitic steel heat exchanger achieves the highest power density at effectiveness values of 0.6 and 0.7, while plastic performs better at higher effectiveness levels. This suggests that advances in thin-wall manufacturing of austenitic steel via AM could make it competitive with, or even superior to, plastic in terms of compactness under certain operating conditions.

% Temporarily sets a new page geometry before the figure:
\newgeometry{top=1cm, bottom=2cm, left=2cm, right=2cm} % Adjust top, bottom, left, and right margins

% Restores the original layout settings after the figure:
\restoregeometry % Restore the default page layout

Broadening the scope to consider the influence of effectiveness values, Figure~\ref{NPD_vs_t_d} also shows that the impact of plate thickness on power density becomes less significant at effectiveness values of 0.9 and above for all materials. This is because, at these high effectiveness levels, the heat exchangers are already dimensionally large, even at the smallest plate thickness considered. As a result, the additional increase in size caused by a thicker plate has a relatively limited impact on the volume of the heat exchanger and, consequently, on the power density.
%%%%
\section{Discussion}\label{sec:discussion}
This study demonstrates that materials with lower thermal conductivity, such as plastics and ceramics, exhibit greater potential for maximizing compactness in cHEX design compared to metals like copper or aluminum. This is primarily attributed to the dominance of axial heat conduction in highly thermally conductive materials, which diminishes effectiveness in cHEXs. Consequently, for a given effectiveness, heat exchangers made from materials with higher thermal conductivity need to be larger than those made from lower thermal conductivity materials. However, the final compactness of a cHEX is influenced not only by material properties but also by manufacturing and fouling constraints, such as the minimum achievable plate thickness and the required minimum plate spacing.

In optimizations conducted without manufacturing or fouling constraints at $\varepsilon_d$~=~0.79, plastic achieves the highest power density, with an improvement factor of 962 relative to the initial heat exchanger. When a uniform plate thickness of 0.5\,mm is applied to all materials, plastic continues to outperform others, reaching an improvement factor of 4.36 compared to the additive manufacturable reference plastic design. 
However, when the fouling constraint is introduced at the same fixed plate thickness, the optimal plastic design exhibits a lower power density than the reference plastic design. Despite this reduction, plastic still performs better than the other materials, followed closely by austenitic steel. In this optimization scenario, the plastic achieves 1.02 times the performance of austenitic steel.

When material-specific plate thicknesses are applied to reflect the resolution limits of different AM processes, the performance advantage of plastic becomes evident once more, with an improvement factor of 1.52 relative to the additive manufacturable reference plastic design. This is primarily because AM thin-wall production has not yet reached the desired precision for metal and ceramic materials. With material-specific plate thicknesses, the power density of the plastic heat exchanger is 7.5 times higher than that of the optimal copper heat exchanger. Similarly, the power density of austenitic steel is 6.06 times higher, Al\textsubscript{2}O\textsubscript{3} is 5.87 times higher, AlN is 2.81 times higher, and aluminum is 2.06 times higher than copper. This means that even the aluminum heat exchanger achieves nearly half the volume of the copper heat exchanger in this optimization scenario. This outcome is driven not only by the increased volume associated with thicker plates but also by the effect of thicker plates enhancing axial heat conduction. This increased conduction necessitates longer heat exchanger lengths and wider plate spacings to achieve the desired design effectiveness.

The observation that the copper heat exchanger has the lowest power density compared to other materials holds true for all $\varepsilon_d$ in the range of 0.55 to 0.94 when using material-specific plate thicknesses, as shown in Figure~\ref{opt_results_varying_eff}. Meanwhile, plastic consistently outperforms all other materials across this range in terms of power density, though other criteria like operational temperatures might also influence the material selection. However, as expected, the power density decreases with rising effectiveness for all materials, indicating the increase in heat exchanger dimensions with effectiveness. The increase in plate spacing with effectiveness can be observed by examining the activation of the fouling constraint in the same figure. At high effectiveness values, particularly for plates with high thermal conductivity, the fouling constraint becomes inactive. This occurs because the maximum power density is achieved with a plate spacing configuration exceeding the 0.8 mm minimum requirement.

These trends can also be understood in light of the asymptotic high-NTU effectiveness limit. For counterflow arrangements this limit is given by 
\[
\varepsilon_{\mathrm{NTU}\rightarrow\infty} = \frac{\M+1}{2\M+1},
\]
with \(\M\) defined in Eq.~\ref{M}. This expression shows that, independent of channel length, the attainable effectiveness may be capped below the design target if \(\M\) is sufficiently large. The controlling manufacturing property in this regard is the product \(k_w t_{\text{min}}\), i.e., the material thermal conductivity multiplied by the minimum manufacturable wall thickness. For high-conductivity materials, AM processes generally require larger wall thicknesses, and this combination restricts the effectiveness achievable in compact designs. While such limitations could be alleviated by allowing for higher pressure drops or larger plate spacings, the latter comes at the expense of compactness and illustrates an inherent trade-off that makes very high conductivity materials less suitable for small flow passages in additively manufactured heat exchangers. These trade-offs further emphasize that plate thickness plays a crucial role in determining the maximum achievable power density, as pointed out in Section~\ref{sec:results3}.

The optimizations in Section~\ref{sec:results3} evaluate how advancements in AM resolution could enhance power density or influence the optimal material selection for cHEX designs.
Figure~\ref{NPD_vs_t_d} demonstrates that the limitations of AM in producing thin features constrain the maximum power density. Reducing plate thickness increases the maximum power density across all examined materials, with the effect being most pronounced for copper compared to plastic and austenitic steel. These results highlight the potential for advancements in thin-wall manufacturing to significantly improve the compactness of copper heat exchangers. Despite this potential, copper’s high thermal conductivity still results in larger heat exchangers compared to austenitic steel and plastic. Even with plate thicknesses of $t^\ast=0.63$ ($t = 0.1\,\text{mm}$) or $t^\ast=1.25$ ($t = 0.2\,\text{mm}$), copper consistently achieves the lowest power density across all effectiveness values.

Interestingly, when the plate thickness is set to 0.1\,mm or 0.2\,mm, austenitic steel outperforms other materials in power density for effectiveness values of 0.6 and 0.7. These results align with previous findings in the literature. Stief et al. High thermal conductivity in cHEX plates can impair thermal performance due to increased axial conduction, while excessively low conductivity may lead to insulation-like behavior, also diminishing performance~\cite{Stief1999}. At an effectiveness of 0.6, the power density of austenitic steel exceeds that of plastic by factors of 1.02 and 1.04 for plate thicknesses of 0.1\,mm and 0.2\,mm, respectively. However, as the target effectiveness increases to 0.79, the power densities of plastic and austenitic steel converge and become nearly identical for both thickness values. At higher effectiveness levels, plastic consistently outperforms austenitic steel. While these findings highlight the potential of austenitic steel in specific low-effectiveness scenarios, the thin plate values required for such performance still exceed current AM capabilities for producing structurally robust, leak-tight steel heat exchangers.

This study focuses on a counterflow PFPHE. However, the results obtained for maximum power density without plate thickness and fouling constraints are also compared with those of a crossflow PFPHE. The analysis of crossflow PFPHE utilized the correlations presented by Buckinx et al.~\cite{Buckinx2013}. For the crossflow heat exchangers, the same design effectiveness value of 0.79 was used, and the maximum power density, along with the corresponding optimal length and plate spacing for the same six materials, was calculated. The results showed that the inverse relationship between maximum power density and thermal conductivity observed in the counterflow case was maintained. However, for each material, crossflow heat exchangers exhibited approximately nine times lower maximum power density compared to counterflow heat exchangers.

Finally, it is crucial to note that although plastic yields the smallest volume heat exchanger for most of the optimization scenarios, its susceptibility to high temperatures might limit practical applicability.

%%%
\section{Conclusions}\label{sec:conclusion}
This study investigates the influence of additive manufacturing (AM) and fouling constraints on the sizing of compact heat exchangers (cHEXs) for different materials. The analysis is based on a nondimensional thermal model that captures the coupled effects of fluid flow, heat transfer, and axial conduction, and its accuracy is validated through CFD simulations, showing close agreement with analytical predictions. This model is applied in numerical optimization to evaluate power density under additive manufacturing and fouling constraints for 6 different materials. 

The findings reveal that, for a given effectiveness, materials such as a representative low-conductivity plastic, austenitic steel, copper, aluminum, aluminum nitride (AlN), and aluminum oxide (Al\textsubscript{2}O\textsubscript{3}) enable varying levels of compactness. Among these, plastic consistently enables more compact heat exchanger designs in the majority of optimization scenarios. The greatest performance advantage occurs when manufacturing and fouling constraints are not applied, yielding an improvement factor (IF) of 962 relative to the baseline design. Even when all materials are subject to a fixed plate thickness constraint of 0.5\,mm, the plastic retains the highest power density, achieving 22 times greater performance than the optimal copper heat exchanger.

However, when fouling constraints are not considered, some configurations result in impractically small plate spacings. For example, the spacing can be as narrow as \SI{22}{\micro\meter} in the case without manufacturing or fouling constraints, or \SI{151}{\micro\meter} when only a plate thickness constraint is applied for plastic. These dimensions make the designs highly susceptible to fouling. By introducing a fouling constraint and applying the same manufacturable plate thickness across all materials, the power density of austenitic steel becomes comparable to that of plastic.

Copper, despite its high thermal conductivity, consistently exhibits the lowest power density in all optimization strategies due to significant performance degradation caused by axial conduction in the wall. This results in less compact designs. High thermal conductivity materials are further limited by the relatively lower resolution of AM techniques for producing thin walls. Thicker plate requirements worsen axial conduction effects and cause further increases in heat exchanger dimensions when maintaining constant effectiveness.
Using material-specific plate thicknesses in the optimizations accounts for AM's varying capabilities for thin-wall production in different materials. These optimizations confirm that plastic remains the most promising material, benefiting from the relatively higher resolution of AM plastics. This trend is consistent across all effectiveness values considered, with plastic maintaining the highest power density.

The effect of plate thickness on power density underscores the importance of advanced manufacturing techniques in improving cHEX performance. Reducing plate thickness increases power density across all materials, with copper showing the most significant impact due to its high thermal conductivity and axial conduction losses. While plastic maintains the highest power density with thinner plates, advancements in thin-wall manufacturing could make austenitic steel a competitive choice for compact heat exchangers. At an effectiveness of 0.9 or higher, the influence of plate thickness diminishes, as the larger size required to achieve the desired effectiveness reduces its impact on power density.

This study assumes constant material properties for the fluid and solid, and does not consider the mechanical strength of the plates or allowable operating temperatures. For high-temperature applications, the results suggest that austenitic steel may be a favorable choice. Additionally, for simplicity, only unfinned counterflow parallel flat plate heat exchangers are investigated. Future research should address these factors to draw more general conclusions on the most suitable materials for AM cHEXs, and may also extend the present analysis to other flow arrangements, such as cross-flow configurations.

%%%
\section{Acknowledgment}
This work received funding from VLAIO under project number HBC.2021.0801/IAMHEX and from a KU Leuven starting grant with reference STG/21/016.

\bibliography{main}
\end{document}

\setstretch{1.2}

\maketitle

\section*{1. Introduction}

This document provides supplementary material for the paper titled “\textit{Optimal Sizing and Material Choice for Additively Manufactured Compact Plate Heat Exchangers}.” It includes additional figures and tables that support the results and discussions presented in the main manuscript.

The first section presents results from 3D conjugate heat transfer (CHT) simulations conducted in ANSYS Fluent. These simulations are used to validate the effectiveness predictions of the analytical model by comparing them with high-fidelity numerical results. The second section provides the dimensional properties corresponding to the nondimensional designs presented in Table 2 of the main manuscript, which include initial and optimized heat exchanger configurations without manufacturing and fouling constraints. The third section reports the dimensional equivalents of geometrical parameters and power densities for different optimization strategies discussed in Table 3 of the main manuscript. These strategies consider either manufacturing constraints alone or both manufacturing and fouling constraints for a target design effectiveness of \( \varepsilon_d = 0.79 \).

\vspace{1em}

\section*{2. 3D Conjugate Heat Transfer Simulations}

\begin{figure}[H]
    \centering
    \includegraphics[scale=0.75]{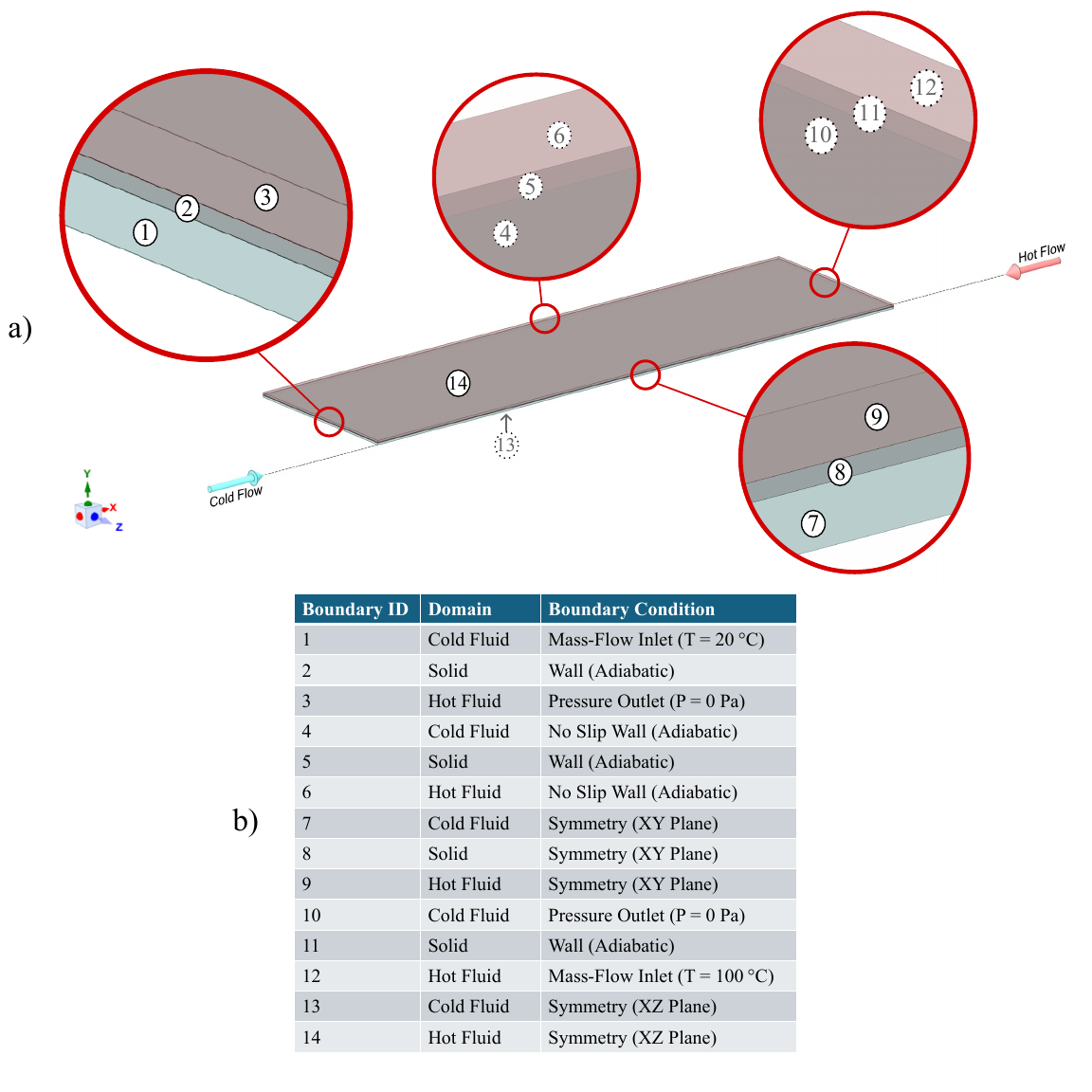}
    \vspace{-4mm}
    \caption{Overview of the 3D conjugate heat transfer simulation setup: (a) computational domain and (b) applied boundary conditions with domain and boundary IDs.}  
    \label{fig:doamin_and_bcs_sup}
\end{figure}

This section presents effectiveness results from 3D conjugate heat transfer (CHT) simulations performed using ANSYS Fluent. These simulations were conducted for selected cases to evaluate the accuracy of the analytical model in predicting the thermal effectiveness of compact plate heat exchanger designs. The computational domain and boundary conditions used in the simulations are illustrated in Figure~\ref{fig:doamin_and_bcs_sup}, and the resulting effectiveness values are summarized in Table~\ref{tab:eff_cfd_sup}.

\vspace{-4mm}
For these simulations, both fluid streams have identical heat capacity rates, resulting in a balanced heat exchanger configuration. Under this condition, thermal effectiveness from CFD, denoted as \( \varepsilon_{\text{CFD}} \), can be determined directly from bulk temperatures without requiring mass flow rate or specific heat values:
\begin{equation}
    \label{eq:eff_cfd}
    \varepsilon_{\text{CFD}} = \frac{T^{\text{b}}_{\text{out,cold}} - T^{\text{b}}_{\text{in,cold}}}{T^{\text{b}}_{\text{in,hot}} - T^{\text{b}}_{\text{in,cold}}}
    = \frac{T^{\text{b}}_{\text{in,hot}} - T^{\text{b}}_{\text{out,hot}}}{T^{\text{b}}_{\text{in,hot}} - T^{\text{b}}_{\text{in,cold}}},
\end{equation}
where \( T^{\text{b}} \) represents the mass-flow-averaged fluid temperature at the corresponding surface. Both expressions result in the same effectiveness within numerical precision due to energy conservation.

To quantify the deviation between CFD and the analytical model, the relative difference (RD) is defined as:
\begin{equation}
    \label{eq:relative_difference}
    \text{RD} = \frac{\left| \varepsilon_{\text{CFD}} - \varepsilon_{\text{model}} \right|}{\varepsilon_{\text{CFD}}}.
\end{equation}
The analytical value \( \varepsilon_{\text{model}} \) is fixed at 0.791 in all cases due to the imposed design constraint. The comparisons show excellent agreement, with all relative differences remaining below 0.52\%.

\begin{table}[!ht]
    \centering
    \caption{Comparison of effectiveness values obtained from CFD simulations and the analytical model for different cases\textsuperscript{*}.}
    \begin{threeparttable}
        \small
        \begin{tabular}{p{4cm} p{3.5cm} c c}
            \toprule
            \textbf{Case} & \textbf{Material} & \(\varepsilon_{\text{CFD}}\) & RD [\%] \\
            \midrule
            Initial (Baseline) & Austenitic steel & 0.7929 & 0.24 \\
            \midrule
            \multirow[c]{3}{4cm}{Additive Manufacturable Reference} 
                & Plastic             & 0.7930 & 0.25 \\
                & Austenitic steel    & 0.7938 & 0.35 \\
                & Ceramic (Al\textsubscript{2}O\textsubscript{3}) & 0.7911 & 0.01 \\
            \midrule
            \multirow[c]{6}{4cm}{Material-Specific Thickness and Fouling Constraints}
                & Plastic             & 0.7951 & 0.52 \\
                & Austenitic steel    & 0.7926 & 0.20 \\
                & Ceramic (Al\textsubscript{2}O\textsubscript{3}) & 0.7925 & 0.18 \\
                & Ceramic (AlN)       & 0.7916 & 0.07 \\
                & Aluminum            & 0.7915 & 0.06 \\
                & Copper              & 0.7912 & 0.02 \\
            \bottomrule
        \end{tabular}
        \begin{tablenotes}
          \footnotesize
          \item[*] The analytical effectiveness value is fixed at \( \varepsilon_{\text{model}} = 0.791 \) in all cases due to the design constraint.
        \end{tablenotes}
    \end{threeparttable}
    \label{tab:eff_cfd_sup}
\end{table}

\vspace{1em}

\section*{3. Dimensional Results for Baseline and Optimized Designs without Manufacturing and Fouling Constraints}

Table~2 of the main manuscript presents optimization results for six materials: copper, aluminum, aluminum nitride (AlN), aluminum oxide (Al\textsubscript{2}O\textsubscript{3}), austenitic steel, and plastic. The objective was to maximize power density without imposing constraints related to manufacturing or fouling. This section provides the dimensional values for the initial (baseline) and optimized heat exchanger designs corresponding to those nondimensional results. Reported quantities include length \( L \), plate spacing \( D \), wall thickness \( t \), channel width \( W \), number of channels \( n \), mass flow rate per unit width for a single channel \( \dot{m}^{\prime}_\text{single} \), and dimensional power density \( \mathcal{D} \). The data are listed in Table~\ref{tab:opt_without_manuf_const_sup}.

All designs in this study assume a fixed total thermal power. As a result, the product \( nW\dot{m}^{\prime}_\text{single} \), which determines the total mass flow rate, is constant across initial and optimized configurations. Since \( n \) and \( W \) are independent variables, multiple combinations can satisfy this condition. In this study, \( W \) was selected such that the width-to-spacing ratio of the optimized designs remains approximately equal to that of the baseline design. This ensures a sufficiently large channel aspect ratio to support the 2D flow assumption used in the analytical model.

\begin{table}[!ht]
    \centering
    \caption{Dimensional parameters for initial (baseline) and optimized heat exchanger designs corresponding to the nondimensional results presented in Table~2 of the main manuscript.}
    \begin{threeparttable}
        \resizebox{0.9\textwidth}{!}{
            \begin{tabular}{
              l
              S[table-format=6.2]
              S[table-format=1.3]
              S[table-format=1.4]
              S[table-format=5.2]
              S[table-format=6.0]
              S[table-format=1.2e1]
              S[table-format=5.2] }
                \toprule
                {\textbf{Material}} & {\textbf{$L$ [mm]}} & {\textbf{$D$ [mm]}} & {\textbf{$t$ [mm]}} & {\textbf{$W$ [mm]}} & {\textbf{n [-]}} & {\textbf{$\dot{m}^{\prime}_\text{single}$ [kg/s·m]}} & \textbf{\( \mathcal{D} / 10^{8}\) [W/m\(^3]\)} \\ 
                \midrule
                \rowcolor{gray!15}
                Austenitic steel (Initial)  & 158.00  & 1.000  & 0.1600 & 95.00 &     40 & 4.75e-3 & 8.26 \\
                Plastic                     &   0.11  & 0.022  & 0.0035 &  2.09 & 121169 & 7.14e-5 & 7964.35 \\
                Austenitic steel            &  10.95  & 0.220  & 0.0352 & 20.90 &   1183 & 7.31e-4 & 83.38 \\
                Ceramic (Al\textsubscript{2}O\textsubscript{3}) & 14.64  & 0.253  & 0.0405 & 24.06 &    902 & 8.33e-4 & 61.75 \\
                Ceramic (AlN)               &  98.49  & 0.660  & 0.1056 & 62.49 &    132 & 2.19e-3 & 9.26 \\
                Aluminum                    & 129.31  & 0.755  & 0.1208 & 71.57 &    101 & 2.50e-3 & 7.03 \\
                Copper                      & 217.80  & 0.981  & 0.1570 & 93.98 &     59 & 3.26e-3 & 4.19 \\       
                \bottomrule
            \end{tabular}
        }
    \end{threeparttable}
    \label{tab:opt_without_manuf_const_sup}
\end{table}

\vspace{1em}

\section*{4. Dimensional Results for Optimized Designs with Manufacturing and Fouling Constraints}

Table~3 of the main manuscript presents optimization results for the same six materials, this time incorporating manufacturing and fouling constraints. These include fixed or material-specific plate thickness limits and a minimum allowable plate spacing to reflect additive manufacturing capabilities and operational fouling limits. This section presents the dimensional equivalents of the nondimensional parameters shown in Table~3. Dimensional plate thickness \( t \), spacing \( D \), length \( L \), width \( W \), the number of channels \( n \), and power density \( \mathcal{D} \) values are listed for each material under multiple optimization strategies, including both reference and optimized designs. As in Section~3, \( n \) and \( W \) are determined based on the fixed total thermal power assumption, such that the product \( nW\dot{m}^{\prime}_\text{single} \) remains constant across all designs. The results are summarized in Table~\ref{tab:opt_with_manuf_const_sup}.

\begin{table}[!h]
    \centering
    \caption{Dimensional geometrical parameters and power densities for different optimization strategies and reference cases with \( \varepsilon_d = 0.79 \) corresponding to the nondimensional results presented in Table~3 of the main manuscript.}
    \begin{threeparttable}
        \small
        \resizebox{0.82\textwidth}{!}{%
        \begin{tabular}{>{\raggedright\small}m{0.26\linewidth}>{\raggedright\small}m{0.20\linewidth}rrrrrrr}
            \toprule
            \textbf{Case} & \textbf{Material} & \textbf{\( t \) [mm]} & \textbf{\( D \) [mm]} & \textbf{\( L \) [mm]} & \textbf{\( W \) [mm]} & \textbf{\( n \) [-]} & \textbf{\( \mathcal{D} / 10^{8} \)} [W/m\(^3\)] \\ 
            \midrule
            \multirow{6}{*}{\thead{\small Uniform Thickness  \\ \small Constraint}} 
            & Plastic            & 0.50 & 0.151 & 5.78 & 14.37 & 2790 & 38.19 \\
            & Austenitic steel   & 0.50 & 0.596 & 70.79 & 56.61 & 142 & 9.22 \\
            & Ceramic (Al\textsubscript{2}O\textsubscript{3}) & 0.50 & 0.684 & 90.47 & 65.27 & 104 & 7.91 \\
            & Ceramic (AlN)      & 0.50 & 1.227 & 302.43 & 115.77 & 34 & 2.79 \\
            & Aluminum           & 0.50 & 1.346 & 363.63 & 128.05 & 28 & 2.38 \\
            & Copper             & 0.50 & 1.594 & 511.53 & 151.96 & 20 & 1.76 \\ 
            \midrule
            \multirow{6}{*}{\thead{\small Uniform Thickness \\ \small and Fouling Constraints}}
            & Plastic            & 0.50 & 0.80\textsuperscript{*} & 108.42 & 76.02 & 67 & 8.00 \\
            & Austenitic steel   & 0.50 & 0.80\textsuperscript{*} & 109.49 & 75.63 & 68 & 7.85 \\
            & Ceramic (Al\textsubscript{2}O\textsubscript{3}) & 0.50 & 0.80\textsuperscript{*} & 113.35 & 76.06 & 70 & 7.32 \\
            & Ceramic (AlN)      & 0.50 & 1.227 & 302.43 & 115.77 & 34 & 2.79 \\
            & Aluminum           & 0.50 & 1.346 & 363.63 & 128.05 & 28 & 2.38 \\
            & Copper             & 0.50 & 1.594 & 511.53 & 151.96 & 20 & 1.76 \\
            \midrule
            \multirow{6}{*}{\thead{\small Material-Specific Thickness \\ \small and Fouling Constraints}} 
            & Plastic            & 0.10 & 0.80\textsuperscript{*} & 101.34 & 75.56 & 63 & 13.26 \\
            & Austenitic steel   & 0.25 & 0.80\textsuperscript{*} & 104.32 & 76.57 & 64 & 10.71 \\
            & Ceramic (Al\textsubscript{2}O\textsubscript{3}) & 0.25 & 0.80\textsuperscript{*} & 106.05 & 75.48 & 66 & 10.37 \\
            & Ceramic (AlN)      & 0.25 & 0.964 & 188.37 & 92.01 & 55 & 4.97 \\
            & Aluminum           & 0.30 & 1.148 & 261.90 & 109.55 & 38 & 3.64 \\
            & Copper             & 0.50 & 1.594 & 511.53 & 151.96 & 20 & 1.76 \\
            \midrule
            \multirow{5}{*}{\thead{\small Additive Manufacturable \\ \small Reference}} 
            & Plastic            & 0.10 & 1.000 & 157.76 & 94.86 & 40 & 8.74 \\
            & Austenitic steel   & 0.25 & 1.000 & 159.31 & 95.79 & 40 & 7.53 \\
            & Ceramic (Al\textsubscript{2}O\textsubscript{3}) & 0.25 & 1.000 & 160.63 & 94.23 & 41 & 7.41 \\
            & Ceramic (AlN)      & 0.25 & 1.000 & 196.34 & 94.44 & 50 & 4.96 \\
            & Aluminum           & 0.30 & 1.000 & 235.16 & 94.26 & 60 & 3.33 \\            
            \bottomrule
        \end{tabular}
        }
        \begin{tablenotes}
            \footnotesize
            \item[*] Fouling constraint is active.
        \end{tablenotes}
    \end{threeparttable}
    \label{tab:opt_with_manuf_const_sup}
\end{table}